# A novel compound synapse using probabilistic spin-orbit-torque switching for MTJ based deep neural networks

Vaibhav Ostwal[1,2], Ramtin Zand[3], Ronald DeMara[4], Joerg Appenzeller[1,2]

[1,2] Electrical and Computer Engineering, Purdue University  [3] Computer Science and Engineering, University of South Carolina  [4] Electrical and Computer Engineering, University of Central Florida

**Abstract:** Analog electronic non-volatile memories mimicking synaptic operations are being explored for the implementation of neuromorphic computing systems. Compound synapses consisting of ensembles of stochastic binary elements are alternatives to analog memory synapses to achieve multilevel memory operation. Among existing binary memory technologies, magnetic tunneling junction (MTJ) based Magnetic Random Access Memory (MRAM) technology has matured to the point of commercialization. More importantly for this work, stochasticity is natural to the MTJ switching physics e.g devices referred as p-bits which mimic binary stochastic neurons. In this article, we experimentally demonstrate a novel compound synapse that uses stochastic spin-orbit torque (SOT) switching of an ensemble of nano-magnets that are located on one shared spin Hall effect (SHE) material channel, i.e. tantalum. By using a properly chosen pulse scheme, we are able to demonstrate linear potentiation and depression in the synapse, as required for many neuromorphic architectures. In addition to this experimental effort, we also performed circuit simulations on an SOT-MRAM based 784×200×10 deep belief network (DBN) consisting of p-bit based neurons and compound synapses. MNIST pattern recognition was used to evaluate the system performance, and our findings indicate that a significant reduction in recognition error rates can be achieved when using our incremental pulse scheme rather than a non-linear potentiation and depression as obtained when employing identical pulses.

**Introduction:** Analog electronic non-volatile memories (eNVMs) have attracted attention in the research community for their potential as synaptic elements [1]. The conductance of such an eNVM can increase or decrease in a continuous analog fashion, mimicking the potentiation or depression of a synapse. However, while Resistive Random Access Memory (RRAM) technology has shown the potential for achieving such analog conductance behavior, the reliable fabrication of analog RRAM devices has remained challenging [2]. Hence, compound synapses that consist of an ensemble of *binary* memory elements have been proposed. Employing the probabilistic switching of individual memory elements, multilevel operation can be realized in a reproducible fashion. In fact, simulations of spiking neural networks (SNN)[3] using compound synapses from binary memory devices can elucidate the desired performance specifications. Moreover, experimental implementations based on an arrangement of parallel binary RRAM devices and simulations of convolutional neural networks (CNNs) [4] demonstrated multi-level operation of such compound synapse structures. However, RRAM technology is facing challenges in terms of current and voltage scaling and is prone to process variability and instabilities. On the other hand "MRAM has already found a niche market and is heading toward disruptive growth" according to Bhatti et al. [5]. Spin transfer torque (STT)-MRAM is close to foundry scale production [6][7] and wafer-scale manufacturability has been shown even for SOT-MRAM [8]. In fact, compound synapses based on a series arrangement of STT-MTJs have already been demonstrated [9][10].

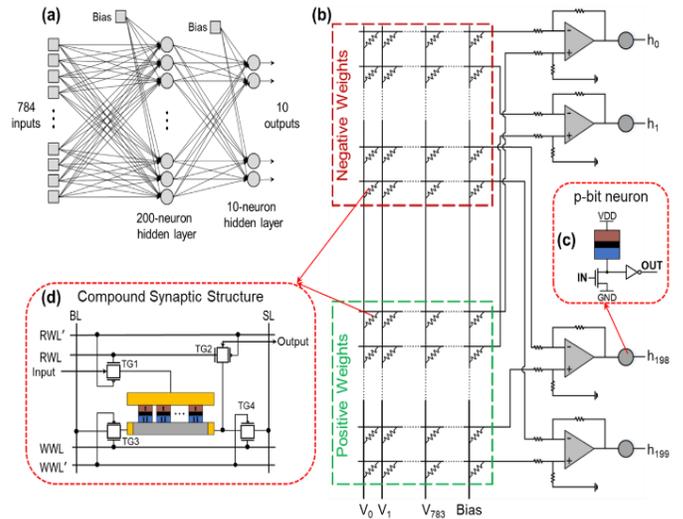

Fig. 1: (a) Graph representation of the 784×200×10 DBN (b) equivalent circuit for the first layer, (c) p-bit as neuron and (d) compound synapse implemented with MRAM cells.

Here we propose a new SOT-MRAM based compound synaptic structure as shown in figure 1(d) as part of a neural network that utilizes yet another MTJ based element, i.e. a p-bit in figure 1(c) [11] as a binary stochastic neuron. (For details on the topic of p-bits see e.g. [11]. While the above mentioned STT-MTJ based synapse shares the same READ and WRITE path, implying that the resistance of the READ path affects the resistance of the WRITE path, our SOT synapse does not suffer from this problem, since writing occurs by means of the spin Hall effect (SHE), as discussed below. Moreover, in general SOT-MRAM is expected to perform better in terms of endurance, power consumption and speed [8]. In this article, we first experimentally demonstrate a novel spintronics compound SOT based synapse (figure 3a) and then utilize a modified version of this device (figure 1d) to simulate a spin-based based deep belief network (DBN). In particular, we have explored the accuracy of a 784×200×10 DBN as shown in figure 1 for MNIST pattern recognition, noting that the realization of a synaptic network with MRAM elements presents an attractive opportunity to build an all-MRAM based DBN. It is also worthwhile noticing that the compound synaptic structure proposed here can readily be combined with other neuromorphic computing architectures as discussed in [10][12][13][14].

For the experimental demonstration we utilize the intrinsic property of spin devices to exhibit thermally activated switching of their magnetization that is probabilistic in nature. We have built and characterized an array of nanomagnets with perpendicular shape anisotropy (PMA) located on a tantalum layer that acts as a spin Hall effect (SHE) channel. Probabilistic switching of the individual nanomagnets is used to realize our SOT synapse. Since an individual nanomagnet has a finite probability of switching for a properly chosen current pulse through the tantalum layer, the ensemble of nanomagnets shows a gradual increase (potentiation) or decrease (depression) in the total magnetization state similar to an analog memory element. While the observed potentiation and depression is non-linear in general due to the stochastic nature of our compound synapse, a modified pulse scheme, as discussed below, can be used to alleviate this issue.

**Experimental results:** A magnetic stack of Ta(5nm)/CoFeB(1nm)/MgO(2nm)/Ta(2nm) with perpendicular magnetic anisotropy (PMA) was deposited on a Si/SiO$_2$ substrate using sputter deposition techniques. In the first e-beam lithography step, a Hall bar was patterned by etching through the entire Ta/CoFeB/MgO/Ta stack until the SiO$_2$ substrate was reached using Ar ion milling. Subsequently, either a single nanomagnet or an ensemble of nanomagnets are patterned on the Hall bars and defined by etching through the top part of the material stack until the bottom Ta layer was reached. Finally, Ti/Au metal pads were deposited using a standard lift-off process, enabling contacts to the devices. To detect the magnetization state of the system, Anomalous Hall Effect (AHE) measurements were performed using a lock-in scheme after quasi-static pulses of 20 to 50 us in width were applied to switch the nanomagnets. Fig 2(a) shows SOT switching of a fabricated Hall bar with a single nanomagnet using current pulses of 50 us width in the presence of an in-plane external field of 20 mT in the current direction. The Hall bar had a width of around 4 um and the nanomagnet on top is elliptical in shape with axes dimensions of 1 and 3 um respectively and the shorter axis in the current direction. Typically, an external magnetic field is required for SOT driven switching of PMA magnets to break the symmetry of the system [15]. However, it should be noted that this is not a strict requirement. In fact, while we are employing this approach here for proof of principle, there

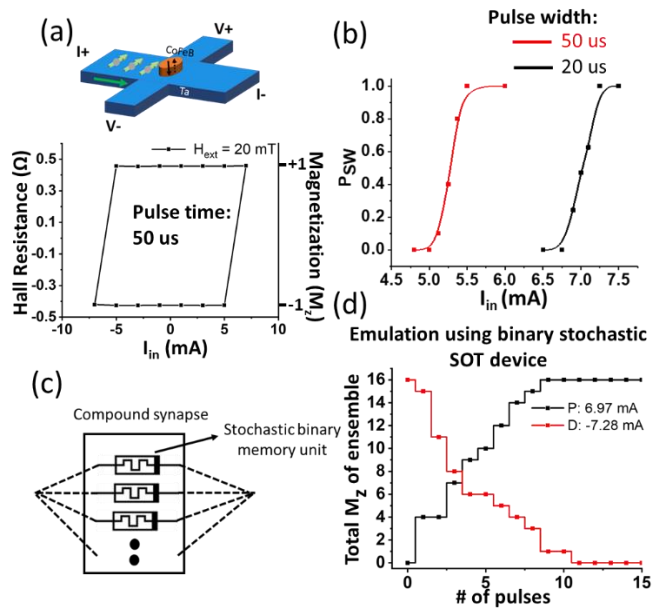

Fig 2 a) SOT switching of a single nanomagnet (binary memory unit) (b) switching probability ($P_{sw}$) curve for two different current pulse widths (c) compound synapse consisting of multiple stochastic binary units (d) experimental potentiation and depression curves emulating a 4-bit compound synapse using an SOT device with a single nanomagnet.

have been demonstrations without external fields that used structures like wedges [16] or novel GSHE materials such as antiferromagnet PtMn [17] to avoid the necessity of using an external field, which is ultimately desirable. To explore the probabilistic nature of SOT switching, we first apply a high negative current (RESET) pulse [width: 50 us] to deterministically switch the nanomagnet to its -1 state, followed by a positive current pulse with varying amplitude (SET) to probabilistically switch the nanomagnet to its +1 state. Fig 2(b) depicts the average probability of switching. As expected, a higher SET current amplitude results in a higher probability of switching. In addition, measurements were also performed employing a pulse width of 20 us, which resulted in a similar trend but higher current requirements for the probabilistic switching, consistent with the expectations for thermally activated spin torque switching. Note that thermally activated switching of nanomagnets is inherently probabilistic in nature and occurs at nanosecond time scales,

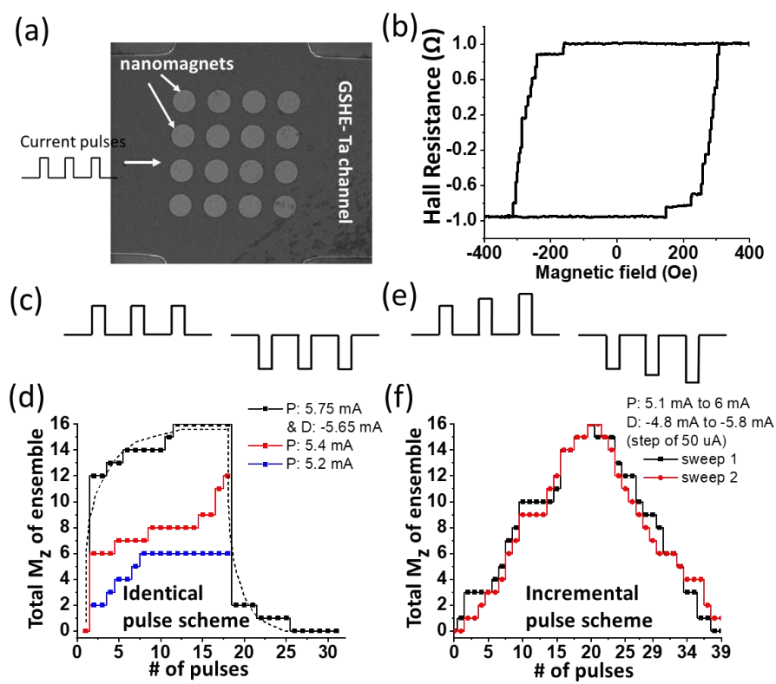

Fig 3 (a) Scanning electron microscope (SEM) image of our SOT based synapse with 16 nanomagnets (b) anomalous Hall effect vs. out-of-plane magnetic field for the device showing 110 mOhm steps, corresponding to individual nanomagnet switching (c) pulse scheme I: identical pulses (d) potentiation and depression curve of the device using pulse scheme I. 3 different potentiation curves showing that the rate of the potentiation can be controlled by the current amplitude (e) pulse scheme II: incremental pulses (f) potentiation and depression curve of the device with pulse scheme II.

which implies that in principle even pulses with widths of nanoseconds are expected to allow for the above described observation.

An n-bit compound synapse can be realized using $2^n$ such nanomagnets operating in the probabilistic switching regime and working in parallel or series (fig 2(c)). To emulate a 4-bit synapse with the "one nanomagnet device", the device was first set to its -1 state as described before. Next, 20 SET pulses at a current level of 6.97 mA with a pulse width of 20 us were applied to the sample. Note that this current level lies in the "unstable region" of the nanomagnet in figure 2(b). Due to the probabilistic nature of switching, the nanomagnet can switch to its +1 state during any of the 20 SET pulses and will remain in this +1 state for all subsequent SET pulses, since the energetic barrier between the -1 and +1 state is of the order of $40k_BT$. Since a 4-bit compound synapse is an ensemble of such 16 devices, our experiment is repeated 16 times on the same device. In this way, instead of using an ensemble of 16 devices, we are able to use a single device to emulate the switching behavior of the ensemble. Fig. 2(d) shows the outcome of this experiment. The number of nanomagnets switched is plotted versus the SET pulse number. The black line clearly illustrates the increasing probability of finding the one nanomagnet switched in all 16 repeated experiments in its +1 state. Such behavior is analogous to the potentiation (P) curve of an analog synapse where the conductance increases gradually with the number of input pulses. Similarly, depression (D) operation can be emulated with our "one nanomagnet device" when input current pulses of -7.28 mA are used – see red curve in fig2(b). Note that by using an MTJ structure instead of our nanomagnet, the plot in fig 2(d) would result in an incremental change in conductance through the MTJ with the number of pulses instead of a change in anomalous Hall resistance, making the approach much more technology relevant and feasible.

Next, we have studied an actual 4-bit compound synapse. A cross shaped Hall bar structure from tantalum with 16 nanomagnets in the cross region was built (see SEM image of the device in figure 3(a)). Fig 3(b) shows the anomalous Hall resistance versus perpendicular external magnetic field of the 16 nanomagnet device with resistance steps that correspond to the switching of individual nanomagnets. Each nanomagnet's switching causes a resistance change of around 110 mOhm ≈ 1.8 Ohm / 16. Due to variations in the coercive fields of the 16 nanomagnets as well as the stochastic nature of the nanomagnets' switching steps occur at slightly different magnetic fields. As in the previous experimental approach, the 16 nanomagnet device is initially put in its -1 state, before being subjected to 20 us pulses of identical amplitude (pulse scheme I). Figure 3(d) shows the resulting P curves for three different current levels in addition to one D curve, displaying very similar characteristics as figure 2(d). As in the "one nanomagnet device" P and D curves of our compound synapse are nonlinear with respect to the pulse number and show a saturating behavior. Note that such non-linear behavior is also observed for analog type RRAM memristors due to the nature of filament formation and breaking during the SET and RESET process. For neural networks, a non-linear response of this type is in general undesirable, since it can have a detrimental impact on the accuracy of the network. To address this issue, researchers [18] have explored a modified pulse scheme in which, instead of using identical pulses, pulses with either incrementally increasing pulse-width or pulse-amplitude are used. Figure 3(f) shows the result of adopting this modified pulse scheme for our 4-bit compound synapse. A much improved linear characteristic is experimentally obtained when increasing the current amplitude from 5 mA to 6 mA (P-curve) and decreasing it from -4.8 mA to -5.8 mA (D-curve) in steps of 0.05 mA (pulse scheme II). Note that for two sweeps (red and black curve in figure 3(f)) the magnetization change is not identical as individual nanomagnets' switching is stochastic in nature. However, the net magnetization of the magnet ensemble shows an overall improved linearity in the characteristics.

**Application-Level Simulations:**

To compare the effect of the two above used switching pulse schemes on the application-level behavior of neuromorphic architectures, we have used the PIN-Sim framework [19] developed by the authors to realize a circuit-level implementation of a deep belief network (DBN) using our compound synaptic structure as weighted connection and embedded MRAM-based p-bits as neurons, as shown in figure 1. In particular, a MATLAB based module [20] is modified according to the characteristics of the neurons to perform an off-line training activity to tune the weights in a 784×200×10 DBN architecture. Once the network is trained, the obtained weights are mapped to various resistive levels that can be realized by our MRAM-based compound synapses, as illustrated by (1):

$$\begin{aligned} &\textbf{\textit{for }} i = 1:(Q+1) \textbf{\textit{ do}} \\ &\textbf{\textit{R}}_{Synapse}(i) = R_P R_{AP}/(R_{AP}(Q-(i-1)) + R_P(i-1)) \\ &\textbf{\textit{end}} \end{aligned} \quad (1)$$

$$R(\theta) = \begin{cases} R_P = R_{MTJ}, & \theta = 0° \\ R_{AP} = R_{MTJ}(1+TMR), & \theta = 180° \end{cases} \quad (2)$$

where the quantization factor ($Q$) defines the number of MTJs used in each compound synapse, and $R_P$ and $R_{AP}$ represent the resistance of the MTJ in its parallel and anti-parallel states, respectively. $R_P$ and $R_{AP}$ are obtained using (2), where $TMR$ is the tunneling magnetoresistance and $R_{MTJ} = RA/Area$, in which $RA$ is the MTJ's resistance-area product, and $Area$ is the surface of the MTJ.

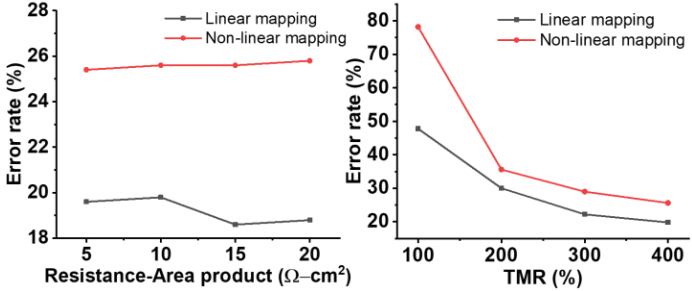

Fig. 4: (a) Error rate versus resistance-area-product, (b) Error rate versus tunneling magneto resistance (TMR) for a 784×200×10 DBN with 4-bit compound synapses.

Figure 1(b) shows the structure of a 784×200×10 DBN circuit implemented by the modified PIN-Sim framework for MNIST pattern recognition application [21]. According to the experimental results shown in figure 3, using the identical pulse scheme (scheme I) leads to simultaneous switching of multiple MTJs in the compound synapses. This results in randomly skipping some of the resistive levels that could be ideally realized by the synaptic structures, which we refer to as non-linear mapping. However, utilization of the incremental pulse scheme (scheme II) enables leveraging all the possible resistive levels that can be realized by a compound synapse, which we refer to as linear mapping. The application-level simulation results for a 4-bit compound synapse (Q=16) are shown in figure 4, which displays the relation between the DBN's error rate and two different device level parameters of the MTJs for nonlinear and linear mapping, respectively. In particular, figure 4(a) focuses on the relation between error rate and $RA$ while the TMR value is assumed to be 400%, while, figure 4(b) shows the error rate versus TMR for RA= 10 $\Omega\mu cm^2$. The obtained results show the significant effect of TMR on the accuracy of the network, while changing the RA value does not result in a major impact on the error rate. Note that a TMR of ~600% has been experimentally demonstrated at room temperature [22]. On the other hand, comparing the impact of the two pulse schemes, a significant decrease in the error rate for the linear mapping method compared to non-linear mapping is observed, which highlights the effectiveness of the incremental pulse scheme approach investigated herein. It is worth noting that reduced error rates can be achieved by using optimization methods such as Genetic Algorithms (GAs) to tune the network parameters such as the network topology, synaptic structure, and sampling mechanism by including device models in the simulation loop. These approaches need to be investigated in future works.

**Conclusion:**

In conclusion, we have experimentally demonstrated proof-of-concept 4-bit compound synapses using probabilistic spin-orbit switching. A modified incremental pulse scheme is shown to result in improved linearity of the synaptic behavior. Furthermore, circuit-level simulations of DBNs consisting of stochastic spin-torque devices, namely p-bits and compound synapses show that the linearity in the synaptic behavior and high TMR-values are crucial device parameters to achieve low error rates.